# Portable Valve-less Peristaltic Micropump Design and Fabrication

H. Yang[1], T.-H. Tsai[2], C.-C. Hu[1]
[1]Institute of Precision Engineering, National Chung Hsing University, Taiwan
[2]Department of Mechanical Engineering, WuFeng Institute of Technology, Taiwan

*Abstract*- This paper is to describe a design and fabrication method for a valve-less peristaltic micro-pump. The valve-less peristaltic micro-pump with three membrane chambers in a serial is actuated by three piezoelectric (PZT) actuators. With the fluidic flow design, liquid in the flow channel is pumped to a constant flow speed ranged from 0.4 to 0.48 mm/s. In term of the maximum flow rate of the micro-pump is about 365μl/min, when the applied voltage is 24V and frequency 50 Hz. Photolithography process was used to fabricate the micro-pump mold. PDMS molding and PDMS bonding method were used to fabricate the micro-channel and actuator chambers. A portable drive controller was designed to control three PZT actuators in a proper sequence to drive the chamber membrane. Then, all parts were integrated into the portable valve-less peristaltic micro-pump system.

## I. INTRODUCTION

Recent progress in MEMS technology provides micro-fluid system manufacture and application pluralism further. Because the micro-fluid systems have the advantages of tiny size and easy to carry, also have high accuracy and short response time. They have quite great values, no matter in such fields like semiconductor, electronic, machinery, chemical analysis or biomedicine and laboratory chip development. Among the micro-fluid control system components, they include micro-channel, micro-valve, micro-pump, micro-sensor and micro-actuator. Micro-pump is the key component in the micro-fluid control system development, since the micro-fluid controls system requires a power to transport fluids via micro-pump. The micro-pump has to control the flow rate accurately and efficiently.

Micro-pump systems usually include an actuating chamber and a valve. There is always a reciprocating cycle vibration in the actuating chamber and the flow channel, it is due to the volume variation of the chamber via membrane moving. A pressure drop is naturally formed when fluids flowing. In accordance with its different actuating model, it can be divided into piezoelectric, electrostatic, thermomechanical, electromagnetic and shape memory alloy types. In addition, the valve can be divided into check-valve and valve-less types. The main function of the valve is to control the flow in a unique direction. The valve of the check-valve type acted as a blocking slice using a cantilever beam in the port. When applied a micro-actuator, it resulted in the difference between the internal and external pressure inside the micro-pump. The valve blocking slice in the channel turns on or off, making the fluid to flow in one way without a reversing flow. Since the valve is operated frequently, it is easy to cause the valve material fatigue or unable to return its original state which affect micro-pump efficiency and life-span. The valve-less micro-pump is comparatively simple and stable.

Koch et al. in 1998 issued the piezoelectric membrane micropump with size of 8mm× 4mm×70μm which deposits PZT material on silicon wafer as actuator [1]. After silicon process of the heat treatment, polarize and electrode, then PZT actuate membrane can be finished, and other structure including actuating chamber, check-valve of cantilever type and entrances-exits are also made with silicon process technology. Micropump have a maximum flow rate 120μl/ min at inputting voltage 600 Vpp with frequency 200 Hz. The advantages of this micropump is lower costs for mass production, the shortcoming is processing complicatedly for PZT membrane. Zengerle et al in 1995 issued the electrostatic type micropump that membrane and operating electrode are separated by insulating barrier [2]. As both ends have a different electric charge which resulted in membrane out of shape with electrostatic, the flow rate is 300μl/ min with maximum moving frequency 1KHz, but it is up to 150~200 operating voltage. Jang in 2000 issued the electromagnetic type micropump, its operating principle is utilizing magnetic and electric field that direction is vertical each other to produce Lorentz force to drive fluids [3]. Its fluid should have conductivity characteristic that can produce Lorentz force and move under the magnetic and electric field reciprocating, it is a kind of non-mechanical micropump. This micropump physical dimension is 40mm×1mm×200μm, the diameter of the entrance and exit of the pipe is 2 mm, when inputting electric current is 1.8 mA, the flow rate is 63μl/ min. Its characteristic is under the interactive influence of the magnetic and electric field for the working fluid to produce Lorentz force and move, so it does not need actuating device on the structural design and does not need to consider the question of the valve either. Its advantage is suitable for high conductivity working fluid, two-way flow, structure simple and process easy; but the shortcoming is that it need plus a regular permanent magnetic field. Stemme issued bimetal micropump in 2000, this micropump membrane is consisted mainly by silicon and aluminum [4]. As the polycrystalline silicon set up an electric circuit and heated, the membrane is crooked and out of shape because the degree of expansion is different when silicon and aluminum are heated. The advantage is its operating voltage low, the shortcoming is the consuming energy and frequency low just only 0.5Hz. 1997, Benard et al., [5] issue memory alloy type micropump, this micropump utilized memory alloy that have different shape characteristics to actuate membrane under different temperature, so it has the function of reciprocating motion micropump under being proper to control, its advantage is difficult to be tired and operate the voltage low for the material, and the shortcoming is operating frequency low and process complicated.

Piezoelectric type micropump which drived by the membrane is developed firstly by Spencer [6] in 1978, micropump combines the piezoelectric actuating slice above the membrane that enable piezoelectric material to drive the membrane to produce high frequency vibration, but after reaching certain vibration frequency, the valve is too late to shut down and cause the phenomenon of back flowing, the main reason is that membrane replying speed can not catch up with its vibration frequency, so the flow rate of micropump is decreasing obviously. Later in 1990, Spencer [7] issue Peristaltic micropump, the major principle is utilizing reciprocating motion of membrane to change the volume of chamber, hence there is a pressure drop inside and outside the chamber that promote fluids to move, its characteristic is three phase place to control the motion of membrane to actuate the glass membrane, resulted in transport function of fluid, frequency at 15 Hz can reach flow rate of 100μl/min, the working figure of micro pump. 1990, Pol [8] issue pneumatic type micropump, one of membrane's end connected to gas pump and squeezed into high-pressure air to make membrane deformation, its advantage is simple process and can verify the experiment fast, but need extra pressure sources is its shortcoming. 2000, Unger [9] choose elasticity PDMS as materials, cooperating with outer way of air pressure to drive membrane to be out of shape, it is mainly two molds of different channel shape after PDMS molding and counterpoint joint, then the atmospheric pressure was pushed into the above channel and then pinches the below fluid of channel to reach the function of pushing the fluid, its channel width is 40μm, height is 10μm, it can be look as the function of valve as one group atmospheric pressure channel, if many group atmospheric pressure channel pinches in a proper order, it can be acted as a Peristaltic micropump, but because atmospheric pressure channel with microchannel overlap area is small, so its maximum flow rate is up to 0.14μl/min only. 1995, Olsson [10] et al., issue diffusing type micropump, its operating principle is to utilize geometry design of the micro-structure, using the difference of come in and go out flow that resulted in diffusion effect to actuate microfluid, its advantage is design and process simple, it is apt to combine it on the chip, and only utilize the geometry form of the micro-structure to cause the function of the valve, so it also call valve-less micropump, but its shortcoming causes the non-directional flow in a chamber of microfluid due to diffusion effect, it is apt to cause the effect of mixing.

The vale-less micro-pump is originated from a diffusing channel concept. It utilizes the micro-structure geometric design by using the difference between inlet and outlet flow that resulted in diffusion effect to actuate the micro-fluid. Its advantages include simple geometry, no active component, providing a constant flow rate. It is apt to combine all components on a chip, and only utilizes the geometrical micro-structures to cause the valve function. That is so call valve-less micro-pump. This paper presents the design and manufacture of the portable valve-less peristaltic micro-pump system, utilizing photolithography technology to produce structural pattern in photoresist and using PDMS material as the molding material. The structure after molding can form flowing channel and actuator chamber, then making a portable circuit controller to control three piezoelectric slices on actuate chamber membranes. By controlling three PZT actuators in a proper sequence, the membrane deformation was actuated directly and generated the wriggling motion in the channel. Due to the valve-less design and the wriggle flow, it is a portable valve-less peristaltic micro-pump system. This paper utilizes the ANSYS to imitate the flow behavior in the micro-channel and results in an optimum solution for the micro-pump.

II. THEORY

Nomenclature
    $A_d$   Narrow Cross-section area of diffuser
    $A_n$   Narrow Cross-section area of nozzle
    $A_{in}$   Cross-section area of entrance
    $A_{out}$   Cross-section area of exit
    Cp   Pressure recovery coefficient
    g   Gravity constant
    K   Loss coefficient
    P   Fluid pressure
    V   Fluid Velocity
    Z   Height of streamline
    $\rho$   Fluid density
    $\xi_d$   Pressure loss coefficient of diffuser
    $\xi_n$   Pressure loss coefficient of nozzle
    $\Phi_d$   Volume flow rate of diffuser
    $\Phi_n$   Volume flow rate of nozzle

Valve-less micropump uses geometry structure of diffuser/nozzle to replace cantilever beam structure of valve to limit the direction of the fluid flow; diffuser/nozzle utilize its difference of conical mouth degree and pressure drop of chamber causing by actuating membrane that make working fluid in diffuser/nozzle have the difference of flow rate and reach the transportation function of the fluid.

It is mainly the pressure recovery coefficient Cp to decide the design of diffuser/nozzle, the relationship figure is the size of diffuser/nozzle and pressure recovery coefficient Cp, the higher is pressure recovery coefficient Cp value, the well design for expand and contract mouth, the lose coefficient K is correlated with pressure recovery coefficient Cp, the larger the pressure recovery coefficient, the smaller loss coefficient K, such as Eq.(1) shown

$$K = 1 - \left(\frac{A_{in}}{A_{out}}\right)^2 - Cp \qquad (1)$$

$A_{in}$ : Cross-section area of entrance
$A_{out}$ : Cross-section area of exit

In order to design better expansion and contraction mouth, it can be got by the relation between K and Cp, it is good design that opening angle 2θ is small, in 5°~20° lose coefficient is minimum, 20°~40° lose coefficient is maximum. According to Bernoulli equation, for any two cross-sectional in pipe for incompressible flow is shown as Eq. (2):

$$\frac{P_1}{\rho g} + \frac{V_1^2}{2g} + Z_1 = \frac{P_2}{\rho g} + \frac{V_2^2}{2g} + Z_2 \qquad (2)$$

P: Fluid pressure, V: Fluid Velocity, Z : Height of streamline, $\rho$ : Fluid density, g: Gravity constant.

If the fluid streamline is highly pretty much the same, it can neglect the gravity iterm, then the Bernoulli equation can be written as Eq. (3):

$$\frac{P_1}{\rho} + \frac{V_1^2}{2g} = \frac{P_2}{\rho} + \frac{V_2^2}{2g} \qquad (3)$$

The structure of diffuser mouth d, it is the increasing cross section that can reduce the velocity of fluid in order to increase the pressure; The structure of nozzle n, it is the cross section decreasing gradually that can increase the velocity of fluid in order to reduce the pressure. As the fluid flows through the structure of the diffuser and nozzle, the pressures can be expressed respectively as Eqs. (4) and (5):

$$\Delta P_d = \xi_d \frac{\rho V_d^2}{2} \qquad (4)$$

$$\Delta P_n = \xi_n \frac{\rho V_n^2}{2} \qquad (5)$$

$\xi_d$ : Pressure loss coefficient of diffuser
$\xi_n$ : Pressure loss coefficient of nozzle

$\xi_d$、$\xi_n$ will change with different channel length、scale and conical degree. By the continuity equation, the flow rate of fluid flowing through the diffuser and nozzle structure can be written as Eqs. (6) (7):

$$\Phi_d = A_d \times V_d \qquad (6)$$
$$\Phi_n = A_n \times V_n \qquad (7)$$

$A_d$ : Narrow Cross-section area of diffuser
$A_n$ : Narrow Cross-section area of nozzle

If the designed structure of diffuser and nozzle is same in size, namely the cross section $A_d = A_n = A$, the volume flow rate can be shown as Eqs. (8) (9):

$$\Phi_d = A \times V_d \qquad (8)$$
$$\Phi_n = A \times V_n \qquad (9)$$

Then the narrow mouth velocity of diffuser and nozzle structure can be shown as respectively Eq.(10)(11):

$$V_d = \sqrt{\frac{2\Delta P_d}{\rho \xi_d}} \qquad (10)$$

$$V_n = \sqrt{\frac{2\Delta P_n}{\rho \xi_n}} \qquad (11)$$

Substituting Eq. (10)(11) into Eq. (8)(9), the volume flow rate of the diffuser and nozzle structure is shown below:

$$\Phi_d = A\sqrt{\frac{2\Delta P_d}{\rho \xi_d}} \qquad (12)$$

$$\Phi_n = A\sqrt{\frac{2\Delta P_n}{\rho \xi_n}} \qquad (13)$$

When the pressure loses coefficient is larger, then the flow rate of volume is relatively smaller; On the contrary, when the pressure loses coefficient is smaller, then the flow rate of volume is relatively larger, hence by good design of the channel length、scale and conical degree to reach the idea flow rate.

### III. EXPERIMENTS

The purpose of this research is to design and fabricate simple structure, easy process, low cost and efficient well for portable valve-less peristaltic micropump system, the major experiment course can be divided into four parts. First is mainly focus on the design of micropump working structure, this research selects valve-less type micropump structure. Second is utilizing ANSYS analyses software to simulate the diffuser structure and nozzle profile to analysis the influence of various angles and lengths of the diffuser and nozzle profiles to get the most stable and efficiency of the physical dimension. Third is to fabricate a microchannel and an actuator chamber. Photolithography process in MEMS technology was used to fabricate main structure and combine PDMS technology to duplicate the microchannel and actuator chamber. Forth is to fabricate the portable control circuit board that can actuate piezoelectric slice and drive the chamber membrane for fluid flowing.

The actuator chamber and channel size of micropump are obtained by according to piezoelectric slice dimension and ANSYS analysis; The structure is shown as Fig. 1, which design three groups of actuator chambers, the size is 10 mm in diameter, another is about diffuser and nozzle geometry size that its key size can be divided several parts. First is entrance width W1, it is in size 300μm, second is diffuser and nozzle mouth angle 2θ that angle is 10 degrees, third is diffuser and nozzle length L that size is 3.4 mm. Analysis of the diffuser is mainly to analyses the influence of the fluid velocity efficiency that the angle causes. The angle 2θ is to be better for smaller, when opening angle 5~20 degrees have minimum of lose coefficient, so selecting simulation angle definitely as 5, 10, 15, 20 degrees of angles, the width W1 of entrance size is 300μm and the length L of the diffuser is 3.4mm, the simulating measurement of the diffuser for different angles is showing as Table 1.

Table 1 Simulation of different angles for the diffuser.

| No | W1(mm) | L(mm) | 2θ  | W2(mm) | L/W1  |
|----|--------|-------|-----|--------|-------|
| 1  | 0.3    | 3.4   | 5°  | 0.597  | 11.33 |
| 2  | 0.3    | 3.4   | 10° | 0.895  | 11.33 |
| 3  | 0.3    | 3.4   | 15° | 1.195  | 11.33 |
| 4  | 0.3    | 3.4   | 20° | 1.5    | 11.33 |

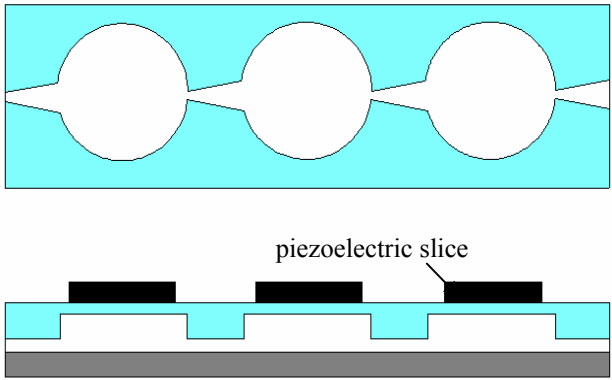

Fig. 1  Schematic diagram of the portable valve-less peristaltic micropump.

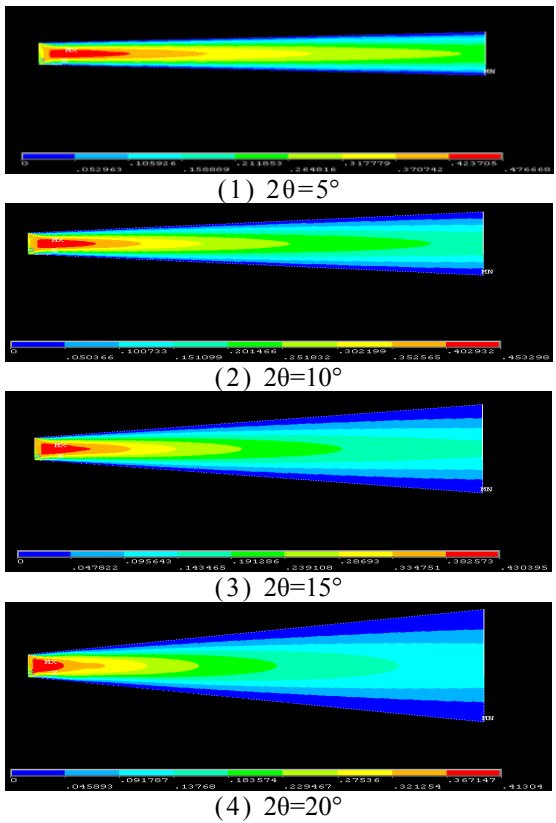

(1)  2θ=5°

(2)  2θ=10°

(3)  2θ=15°

(4)  2θ=20°

Fig. 2  Simulation results of the fluid velocity using different angles for the diffuser.

Fig. 2 is the fluid velocity simulation result from ANSYS for different angle diffuser. Table 2 is the result of imitation for the velocity of flow of the diffuser, from the result of imitating, the width W1 of the entrance of the diffuser is 300μm, length L is 3.4mm, and angle 2θ is 10 degrees that its efficiency is the most stable as shown in Fig. 3. It is the relationship of the opening angle and loss rate of the flow speed; it did not discuss the back flow phenomenon of the diffuser, but the result in simulation is in accord with the theory that is designed, in L/W =11.33 and 2θ=10 degrees, it is one transient stall, its pressure loss is minimum, it is the best performance.

This research utilizes photolithography technology to fabricate a resist structure which include microchannel and actuator chamber and combine PDMS molding technology to duplicate out the microchannel, actuator chamber and membrane as illustrate in Fig. 1. It is a micropump sketching figure; With simple and convenient joint technology, the PDMS molding was bonded on the glass and then use piezoelectric slice to actuate the chamber membrane which utilize vibrating principles of upper and lower piezoelectric slice to drive the membrane out of shape, hence driving the fluids of the chamber to move. This experimental design for driving the chamber is 10mm, using designed actuating circuit to control piezoelectric slice that drive the chamber membrane to reach the function of wriggling. The process of this experiment can be divided into two major steps, first part is utilizing photolithography technology to fabricate a resist structure. The second part is molding, using molding principle to fabricate main structure of micropump by using elastic high polymer material PDMS.

This research uses the JSR-THB negative resist to fabricate JSR mold on the silicon wafer, it is fabricated by the multi-layer coating technology for high aspect ratio microchannel protruding structure. As Fig. 4(a) showing, it is JSR mold finished product figure, it is quite intact that can be observed from actuate chamber and microchannel structure, they did not appear fracture and stripped situation caused by stress concentration. In the course of processing, in order to remove the surplus of the resist solvents, it utilizes baking hard for a long time at low temperature, for avoiding PDMS colloidal to response when PDMS molding appeared unsolidification and stuck on mold; As Fig. 4(b) showing, the chamber and microchannel of PDMS structure, it can be found out the structure to be quite intact.

Table 2 Result of velocity simulation for diffusers.

| No | Inlet(mm/s) | Vmax(mm/s) | Vout(mm/s) |
|----|-------------|------------|------------|
| 1  | 0.35        | 0.48       | 0.261      |
| 2  | 0.35        | 0.45       | 0.192      |
| 3  | 0.35        | 0.43       | 0.153      |
| 4  | 0.35        | 0.41       | 0.120      |

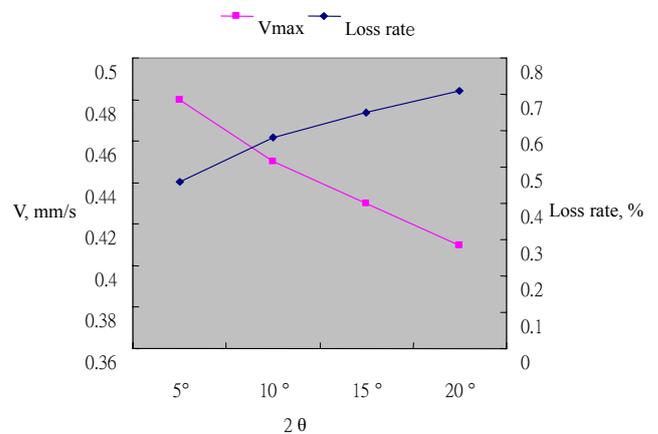

Fig. 3  Relationships of opening angles related to velocity and loss rate.

This resist structure is measured with the 3D confocal microscopy, it utilizes this microscope to take the 3D surface outline of the structure and measures the thickness of the structure. Fig. 5 is the 3D surface outline picture of microchannel, it can find out that microchannel surface is very smooth by using photography process. Fig. 6 is the thickness measurement of the channel, the thickness of channel is nearly 350μm. During the past studying process, other microchannel structures were made. Their thickness ranges were in 346~403μm, it may be an error caused by experimental equipment or the used of negative resist material.

As Fig. 7 showing, this experiment is according to the design step of assembling, it takes with actuator controller, piezoelectric slice and micropump structure to assemble and combine to a portable valve-less peristaltic micropump system, because it did not consider the fluid chamber room when design the micropump originally, so transport fluid in testing, prior to entrance dripping 15ml red dyestuffs to aqueous solution to observe whether the red dyestuffs aqueous solution can flow smoothly in microchannel flow as shown in Fig. 8 (a) (b). The result of the test is that the red dyestuffs aqueous solution can be output smoothly and have not overflowing situation, it is indicated that the microchannel joint strength intensity is enough and have no stopped up phenomenon using PDMS to be joint; Because it did not consider the fluid chamber room, the test of flow rate is measured by sending the volume of the fluid with different frequency, then divided by time, this part is the research that structure need to improve again; As Fig. 9 showing, the flow rate have examined at direct current voltage 24volt for every frequency, but at frequency 50Hz under in a proper order control, the flow rate of volume is about 365μl/min for maximum, according with originally expectation that the flow rate of volume can reach the demand above 200μl/min.

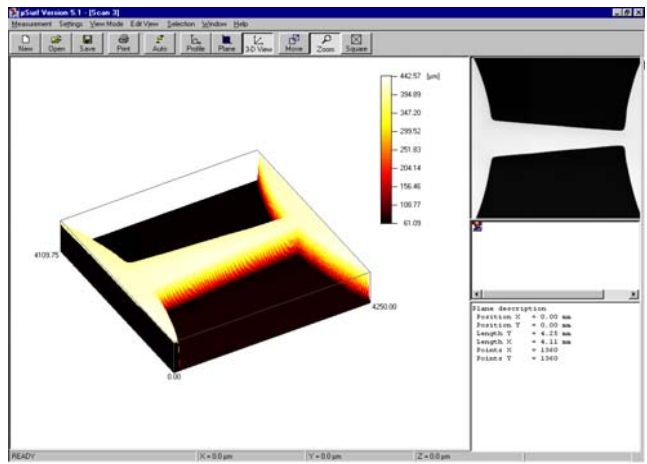

Fig. 5 Three dimensional profile measurement of the flow channel.

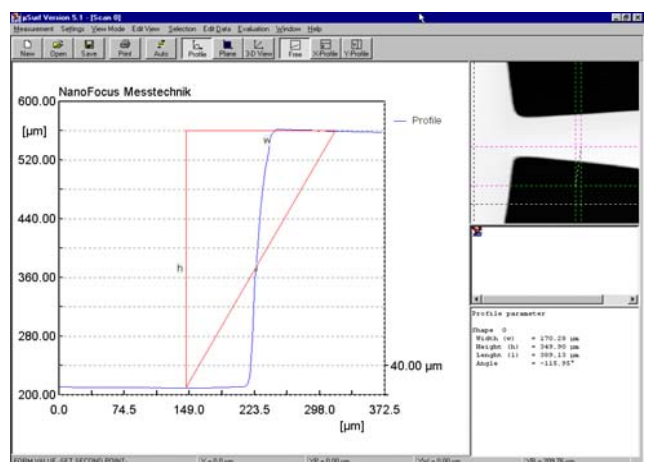

Fig. 6 Thickness measurement of the flow channel.

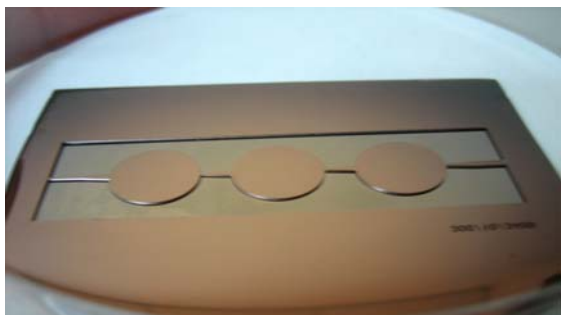
(a)

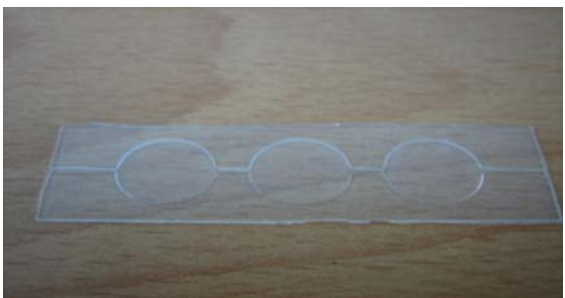
(b)

Fig. 4 Fabrication results of the nozzle and chamber; (a) resist mold in JSR and (b) replicated structure in PDMS.

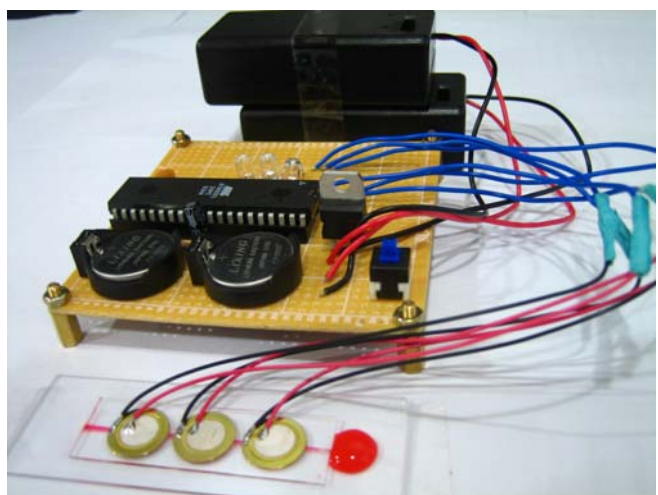

Fig. 7 System assembly of the valve-less peristaltic micropump.

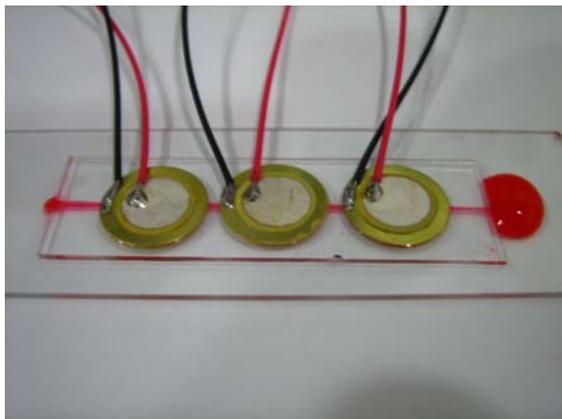

(a)

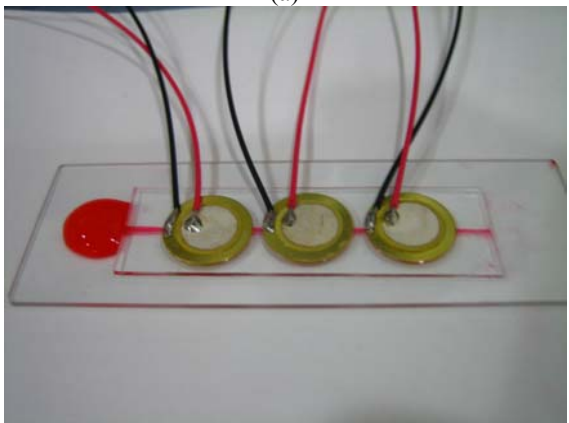

(b).

Fig. 8 Pumping flow test of the valve-less peristaltic micropump; (a) before pumping and (b) after pmuping.

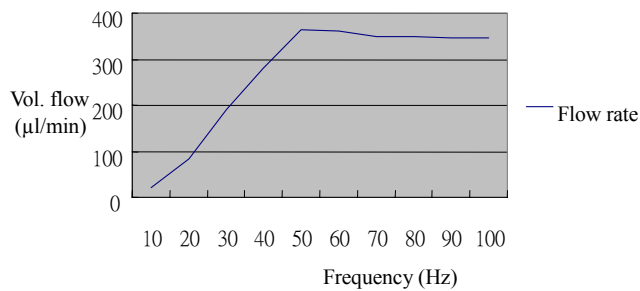

Fig. 9 Relations of volume flow rate related to the operational frequency.

## IV   CONCLUSION

This research is successful using the multi-layer coating method for ultra-thick negative JSR-THB resist to fabricate actuator chamber and microchannel JSR mold of high aspect ratio structure. The characteristics of this research is the micropump structural design and multi-layer coating processing that can reduce process time and cost, and with the portable driving circuit for piezoelectric slices to cooperate with the operational frequency to make three groups of chamber membrane producing wriggles in a progressive way to achieve the goal of promoting fluids. It is obtained the following conclusions from the above experiments.

(1) PDMS molding and PDMS bonding method are used to reduce the process time of photography process. It can reduce cost in mass production.
(2) The processing technique is skillful to control the dimension of the microchannel steadily.
(3) Micropump structure is PDMS that have advantages of elastic and bearing falling etc.
(4) The wriggling transportation of the fluid is a very smooth flow rate and provides a high accurate flow rate.

ACKNOWLEDGMENT

This research is supported by the National Science Council of Taiwan under the grand number NSC95-2212-E-005-091.